# Novel structural evolution of several nanolaminate $M_{n+1}AX_n$ (n=1, 2, 3, *etc.*) ceramics under pressure from first principles

Ben-Yang Li[1], Fang Chen[2], Heng-Na Xiong[1], Ling Tang[1], Ju-Xiang Shao[3], Ze-Jin Yang[1,¯]

[1]School of Science, Zhejiang University of Technology, Hangzhou, 310023, China
[2]School of Chemical Engineering and Technology, North University of China, Taiyuan, 030051, China
[3]Department of Physics, Yibin University, Yibin, 644000, China

**Abstract:**

We did extensive research for the typical nanolaminate $M_{n+1}AX_n$ (n=1, 2, 3, *etc.*) ceramics focusing on the structural stability, the phase transition pressure of $Ti_2GaN$ (~160 GPa) is far higher than that of $Zr_2GaN$ (~92 GPa), meaning the strong "M" dependence of the same group, whereas $Zr_2AlN$ (~98 GPa) has similar value with that of $Zr_2GaN$, meaning the weak "A" dependence. In comparison with the stable $Mo_2GaN$, $Mo_2GaC$ shows lowest phase transition pressure to date among all of the known MAX, meaning that C-containing phase has lower phase transition pressure than that of N-containing counterparts, agreeing with the previously calculated general conclusion that the formation of energy of N-containing MA"N" is twice lower that that of MA"C" counterparts, also meaning the stronger binding strength of N-containing MAX. All of the metastable phases of the selected MAX transition almost at the same time, such as all of the metastable phases of $Zr_2AlN$ transition at the similar pressure, about 90~110 GPa, with a very narrow pressure range of less than 20 GPa, as is also the case for $Zr_2GaN$ corresponding to 90~115 GPa and for

[¯]zejinyang@zjut.edu.cn

Mo$_2$GaC corresponding to 10~25 GPa. Mo$_2$GaC presents multi-phase co-existence status at high pressure, whose *P6$_3$/mmc* (194, *α, β*) *α→β* phase transition is the more commonly and frequently occurred route in this kind of MAX structure. The hexagonal *P6$_3$/mmc* to tetragonal *P4/mmm* transition is the common direct route for the N-containing MAX due probably to the high transition pressure. *P6$_3$/mmc* (194, *α, β*) *α→β→P4/mmm* transition might be the common route for the C-containing MAX due probably to the low *α→β* transition pressure. Nb$_2$InN/Nb$_2$GaN and Mo$_2$InN present *c*-axis abnormal elongation at low pressures, these compounds have negative formation of energy at ambient conditions, meaning that all of them are stable or experimentally synthesizeable.

I. Introduction

The M$_{n+1}$AX$_n$ (n=1, 2, 3, etc.) phase is a class of hexagonal compound, where M is an early transition metal, A is a group 13–16 element, and X is C or/and N[1,2]. These M$_{n+1}$AX$_n$ phases collaboratively combine many excellent properties from metal and ceramics. Mo$_2$GaC, first synthesized in 1967[3], a detailed summary of its progress is reported recently[4], interesting, this paper unexpectedly observed many *c*-axis abnormal elongation behaviours within 0~100 GPa, whereas the structure is always stable below 100 GPa according to their elastic constants and lattice dynamics. Contrarily, a recent study observed the structural phase transition phenomena according to their enthalpy difference[5]. Such similar *c*-axis ultraincompressibility behaviour is also observed in Zr$_2$InC[6]. The structural phase transition is predicted and determined previously in Mo$_2$Ga$_2$C[7,8]. Thus, more similar interesting phenomena might exist, it is necessary to further reveal these phenomena and conclude their common feature. We systematically searched the possible structural phase transition based on these known facts.

## II. Computational methods

The lattice relaxation is done by Vanderbilt-type[9] ultrasoft pseudopotential with Generalized Gradient Approximation (GGA) for the Perdew–Burke–Ernzerhof[10] (PBE) exchange-correlation function, for the calculation consistency and comparison purpose, we used the unified and default ultrafine precision parameter, the energy cut-off is set at 320eV (and evern higher for enthalpy and phonon calculation), the 10×10×2 (and even denser for phonon calculation) Monkhorst-Pack mesh is used in $k$-point sampling[11], the electronic energy convergence and atomic force are set at $0.5×10^{-7}$ and 0.01 eV/Å. The calculation is done by CASTEP[12].

## III. Results and discussion

### A. Phase transition from enthalpy difference: Zr$_2$AlN, Zr$_2$GaN, Ti$_2$GaN, Mo$_2$GaC

A recent high-throuphput structural design of M$_2$AX (211) compounds observed many stable phases under ambient conditions[13] through calculation the formation enthalpy, for simplicity, it is therefore not necessary to repeat the same process, we use their results directly. Moreover, we also ignored the structural information as all of the present lattice structures are same with those of experimental reports[5], in which their computed $c$ axis shows normal contraction at about 15 GPa, differing with another calculation[4]. The structural evolution and enthalpy difference are calculated for Zr$_2$AlN between 80~200 GPa, as are shown in Figures **1,2**, respectively, in which the subtle enthalpy difference at around 100 GPa might also depends on the used function such as LDA (localized density approximation, LDA). The reason why *Cm* phase presenting high enthalpy is that its Zr atom is fixed during relaxation otherwise it will automatically transformed into *C2/m* under ultrafine calculation, *Cm* will remain its symmetry unchanged when the fine or even lower precision calculation is used. The same case is also used in Zr$_2$GaN. The *Cm* and *C2/m* unit cell parameters under 100 GPa are $a=b=2.7757$ Å, $c=20.1971$ Å, $α=β=99.5661°$, $γ=62.8053°$, $V=135.7564$ Å$^3$, $a=b=2.7958$ Å, $c=20.3319$ Å, $α=β=101.609°$, $γ=60.5973°$, $V=134.6438$ Å$^3$, respectively, reflecting the lower enthalpy of *C2/m*. The obvious enthalpy difference could be found above 100 GPa, where the *P4/mmm, C2/m, β* phase are the most, next, second stable structures, the *α→β* phase transition is usually occurred, such as in Ti$_3$SiC$_2$. A similar structural phase transition is calculated[5] for Mo$_2$GaC below 30 GPa, which also shows the general *α→β* phase transition with a negligible exception of *Amm2*, whereas they conclude that the *C2/m* is the most stable phase above 15 GPa, and *C2/c* mixed with the hexagonal phase agrees well the XRD pattern, differing again with their enthalpy difference calculation. The current calculation demonstrates the *β→C2/m* phase transition occurs above about 110 GPa in Zr$_2$AlN, since then the *C2/m* and *β* phases always have the next and second lowest enthalpy until 200 GPa. However, the *C2/m* phase seems couldn't transform to the lowest energy phase in Mo$_2$GaC, as is shown in Figure **3**, in which the inserted small figure also observed the *β→P4/mmm* transition at about 200 GPa. The current phonon dispersion calculation for $Zr_2AlN$ obtained the imaginary frequency at 140 GPa, as is shown in Figure **4**, in accordance with the elastic constants calculation, as is shown in Figure **S1**. The current enthalpy difference calculation of $Zr_2AlN$ finds the *α→β* phase transition at 90 GPa. Phonon vibrations of Zr and Al co-induced the structural instability, as is shown in Figure **S2**, which also illustrates that the Zr-Al bond is easily exploitable to form quasi-plane two-dimensional layered structure. The enthalpy difference per atom is about 10 meV at 100 GPa between *α* and *β* phases, whereas 10 meV is observed in $Mo_2GaC$ only at 30 GPa. Atomic average volume of $Zr_2AlN$ shown in Figure **1** (d) indicates that *α*, *β*, *P4/mmm* phase decreased suddenly with pressure, as is also the case in $Zr_2GaN$ shown in Figure **5** (d). Symbols I, II, III, IV ……in these figures mean the most stable phase at corresponding pressure range, similarly hereinafter.

The structural evolution of $Zr_2GaN$ is shown in Figure **5**, in which the *a/c* axes show the nearly same contraction with a value of about 0.945, meaning the decreased extent of anisotropic degree. $Zr_2GaN$ shows the *α→P4/mmm* phase transition sequence, as is shown in Figure **6**, elastic constants also observed the structural instability as is shown in Figure **S3**.

The phase transition mechanism of *α→β* in MAX is that the "A" shows a relative shift as a whole within the *a,b* projection plane, as is shown in Figure **2**, the *β→C2/m* phase transition is simple as only the lattice *ab* plane transformed from parallelogram of *β* phase to rectangle of *C2/m* phase, both phases have totally same projections towards *c* axis, namely, Zr sites at the center of the Ga/N triangle shape, meaning the almost invariance of the atomic arrangement along *c* axis. Such evolution is controlled by the relatively short and strong N-Zr bond, corresponding to the large phase transition pressure, which keeps the lattice framework unchanged during the compression. The *α,β→Cm,C2/m* phase transition shows even lager lattice deformation and atomic displacement, resulting mainly from the strong "M"-"X" but weak "M"-"A" bond, which is evidenced by the slower "M"-"X" but faster "M"-"A" bond contraction, further making possible the chemical exploitation into quasi-two-dimensional structure. *β→P4/mmm* phase transition has even larger lattice distortion as the Zr-Zr bond has rotated to the *c*-axis orientation. *P4/mmm* phase is a simple structure as it has one formula per unit cell.

The enthalpy difference of $Ti_2GaN$ is shown in Figures **7**, the phase transition pressures of $Ti_2GaN$ (~160 GPa), $Zr_2AlN$ (~98 GPa), and $Zr_2GaN$ (~92 GPa) are higher than those of $Mo_2GaC$ (~10 GPa), $Mo_2Ga_2C$ (~22.3 GPa), indicating the extreme stability and stronger bond strength of N-containing MAX than those of C-containing counterparts despite that the transition is dominated by the shift of "A" other than the "X"(=N) in the most cases, occasionally including the "M", which at least is applicable for the case of *α→β* transition, also revealing that the stronger "M"-"A" bond despite that the "A" is not directly bonded to N.

$Mo_2GaC$ (*P6$_3$/mmc*, 194, *α* phase) stables in a large pressure range of 0~100GPa according to the previous phonon and elastic constants calculation[4]. $Mo_2GaC$ (*P6$_3$/mmc*, 194, *β* phase) is formed above 10 GPa and it has the lowest enthalpy within 10~200 GPa, as is shown in Figure **3**, whose phonon dispersion frequency shown in Figure **S4** is positive value at 30 and 70 GPa, respectively. Calculations for *Amm2* (*αβ*) (No. 38) of $Mo_2GaC$ at 30 and 70 GPa also obtained the positive values, as is shown in Figures **S5,6**. The small enthalpy difference agrees with the low phase transition pressure between *α* and *β* phases, moreover, all of the structural phase transition finished transition simultaneously at about 10~20 GPa, meaning that once the "A" shifts under pressure, its first/second nearest neighbors "M" and "A" remain unchanged until *β* phase is formed. Under the continuous compression, the "M"- "M" bond rotates to parallel the *c* axis orientation, thus the *P4/mmm* phase is formed, which is unique atomic motion behaviour in this kind of layered structure under pressure, as is shown in Figure **2**. The energy difference among these phases are lower than 10 meV/atom at ambient conditions, indicating that these metastable phases could be synthesized experimentally. Previous study[5] suggests the possible phase transition at about 15 GPa based on the XRD measurement and concludes that the possible candidate phase is *C2/c* or *C2/m*, respectively. However the structural phase transition is almost indistinguishable based on the present calculation.

Owing to the *c*-axis abnormal elongation of $Mo_2GaC$ *α* phase from about 15 GPa but *a* axis shows smooth shrinkage behaviour under pressure, thus *c*-axis has sufficient space to allow the Ga atom shift in the basal plane, as is also confirmed by the previous[4] and present calculation for the Mo-Ga length, as is shown in Figure **S7**. As is shown in Figures **S8,9**, the fast shrinkage of *a* axis relative to that of *c* axis forces the Ga shifting towards the geometrical center of the rhombic parallelogram of *a*, *b* basal plane, where also is the symmetry center of the Ga plane and thus has lower energy. Mo-Mo bond of Mo$_2$GaC *α* phase shortens firstly under compression, such process well provide sufficient space to accommodate the incoming Ga atom, meanwhile, it is also the time that the Ga atomic layer shifts towards the center of the *a, b* basal parallelogram plane , This might be the reasonable illustration of the *α→β* phase transition in Mo$_2$GaC and other MAX compounds because this is the most convenient atomic shift route, as is shown in Figure **2**. Such structural phase transition failed to induce the phonon imaginary frequency which could be illustrated by the negligible atomic displacement and energy discrepancy, which might be the true origin of the positive phonon frequency of *β* phase or the other candidate phases, thus confirming the multi-phase stable status of Mo$_2$GaC at high pressure. Such phenomena is frequently occurred in chemical isomers owing to the non-uniquely fixed orientation of the functional groups, such as guanine has dozens of isomers owing to their nearly same energies. Therefore, it is possible for these solid solution MAX phases to have many stable phases at high pressure owing to the unfixed "M" in MAX. Usually, unstable structure also produce positive phonon frequency, the calculated phonon for Mo$_2$GaC (P6$_3$/mmc, 194, *β* phase) at 0 GPa also produced positive frequency, as is shown in Figure **S4**. (Cr$_{0.5}$V$_{0.5}$)$_2$GeC[14] shows elastic metastability at about 600 GPa but the structure is stable as all of the phonon frequency is positive value. Elastic softening behaviour of Ti$_2$GaN[15] induces the structural instability at about 350 GPa that is confirmed by the negative phonon frequency.

Mo$_2$GaC *β* phase shows monotonic axial shrinkage feature, as is shown in Figure **8**, its *c* axis is stiffer than *a* axis. The critical position of *α→β* of Mo$_2$GaC is 10 GPa, the *c*-axis abnormal elongation of Mo$_2$GaC *α* phase appeared also at about 10 GPa, which might imply the potential transient state transformation. Mo$_2$Ga$_2$C will transform its space group from *P6$_3$/mmc* (194, *α* phase) to *P$\bar{3}$m1* (164) at about 22.3 GPa based on the recent experimental and theoretical study[8], and the experiment measurement confirmed a reversible phase transition in Mo$_2$Ga$_2$C[8], meaning a very low energy barrier between the two phases, which also supports evidence for the present analysis of the different phase transitions and confirms the stability of the Mo$_2$GaC *α* phase above 10 GPa. More importantly, these energetic comparable phases at high pressure also challenge the experimental measurement as these phases might be constantly inter-transformation. For simplicity, we only calculated the possible low energy candidate phase with enthalpy no more than about 20 meV per atom relative to the most stable phase at ambient conditions. Usually, metastable phase with energy about 50 meV/atom higher than that of ground state is also possible to be synthesized at some special non-equilibrium technologies. Such as the $E_f$ of Mo$_2$GaC is -0.12811 eV/atom based on previous calculations[13], combined with the present enthalpy calculations of Mo$_2$GaC it is found that the metastable phases have the average energies of about 5 meV/atom at 0 GPa, thus their $E_f$ might be about -0.123 eV/atom, such small energy difference might also be the possible reason of the low transition pressure.

Previous study[5] obtained the highest bulk modulus *B* (GPa) of Mo$_2$GaC *α* phase for all the MAX phase measured to date. The bulk modulus $B$ and its derivative $B'$ of $Mo_2GaC$ $\alpha$ phase is fitted by Birch-Murnaghan and Vinet equation of state within different pressure ranges, respectively, as is shown in Table **S1** (table **S2** is some of the lattice parameters), it is found that the $B$ is distributed at 160~170 GPa with Goodness of fit about 0.999 within 0~100 GPa, which is far smaller than that of 295 GPa measured experimentally[5] within 0~30 GPa. The current result is also evidently smaller than those of results from same function calculated from the elastic constants, such as 189.5 GPa of GGA/PBE without spin[4], 206.2 GPa of GGA/PBE with spin[4], 250.6 GPa of GGA/PBE[16], 191.5 GPa of GGA/PBE[17], 190 GPa of GGA/PW91[18], 248.6 GPa of LDA[19], respectively The large measured $B$ might originate from the mixed compounds other than the unique $\alpha$ phase. The generally fitted $B$ of $\beta$ phase is in the vicinity of 200 GPa, still far smaller than the 295 GPa.

## B. Axial abnormal elongation: $Nb_2InN$, $Nb_2GaN$, $Mo_2InN$,

According to a previous calculation[13] of formation of energy ($E_f$) for many MAX compounds, it is found that the $E_f$ of $Nb_2InN$ should lower than that of $Nb_2InC$ (-0.45504 eV/atom), thus $Nb_2InN$ is stable at ambient conditions. Also, $E_f$ of $Nb_2GaN$ might be comparable or slightly lower than that of $Nb_2InN$. Similarly, the $E_f$ of $Mo_2InN$ is about -0.4 eV/atom based on previous general trend. $Nb_2InN$, $Nb_2GaN$, and $Mo_2InN$ are stable in the pressure range of 0~200 GPa based on previous calculations for $E_f$ at zero pressure and present elastic constants under high pressure shown in Figures **S10-12**, note that three phases of $Mo_2InN$ have nearly same energies at 0 GPa, it is $P6_3/mmc$ (194, $\alpha$ phase), $C2/m$, $Amm2$, respectively, as is shown in Figure **S13**. Nb$_2$GaN will be unstable at about 200 GPa shown in Figure **S12**. The current phonon calculation for Mo$_2$InN at 20 GPa, shown in Figure **S14**, also confirmed its stability, whereas their *c*-axis shows abrupt abnormal elongation and *a*-axis contracts suddenly at low pressures with normal volume shrinkage, within 20~40 GPa in Mo$_2$InN and 5~10 GPa in Nb$_2$GaN and Nb$_2$InN, respectively, as is shown in Figure **9**. A fast *c*-axis abnormal elongation and *a*-axis shrinkage of Mo$_2$InN *α* phase indicates the increased bond strength along *c*-axis but decreased gradually along *a* axis within 20~40 GPa. The enthalpy difference calculations of Mo$_2$InN within 0~80 GPa didn't find structural phase transition behaviour, whereas our calculation for Nb$_2$GaN observed the *β→P4/mmm* transition at about 200 GPa, as is shown in Figure **10**, thus it is reasonable to infer that the Nb$_2$InN and Mo$_2$InN might also exist such phenomena, which might also be the final phase under high pressure of MAX.

Analysis to Nb$_2$GaN, Nb$_2$InN, Mo$_2$InN *α* phase found the common evolution trend, the Nb(Mo)-Nb(Mo) angle rotates deviate the *a*/*b* basal plane quickly, about 5° for Nb$_2$GaN/Nb$_2$InN and 15° for Mo$_2$InN within about 0~40 GPa, respectively, leading to the Nb(Mo)-Nb(Mo) bond elongation and thus resisting the *c*-axis shrinkage, as is shown in Figures **11,12** and Figures **S15** for a larger pressure range. Such rotation becomes very slow above about 40 GPa due mainly to the symmetry constraints during calculation. Nb-Nb bond of Nb$_2$GaN and Nb$_2$InN monotonically elongates for only one time below about 20 GPa but Mo-Mo bond of Mo$_2$InN elongates twice below about 40 GPa, whereas all of the Nb(Mo)-N bonds reduce synchronously in the first time and elongate subsequently, even resulting in the slower shrinkage of *c* axis and faster shrinkage of *a* axis. Therefore, the other MAX compounds might also exist this kind of behaviour.

### C. Axial abnormal elongation: Mo$_2$GaC

Previous $E_f$ calculations[13] for Cr$_2$AlC and Cr$_2$GaC obtained the values of -0.16975 and -0.13176 eV/atom, respectively, the value[13] of Mo$_2$GaC is -0.12811 eV/atom, in addition to the other similar data thus it is feasible to infer that the values of Mo$_2$AlC and Mo$_2$InC might be similar with that of Mo$_2$GaC, in other words, their values should be negative data. In a word, Mo$_2$Ga$_{0.5}$Al$_{0.5}$C and Mo$_2$Ga$_{0.5}$In$_{0.5}$C also might have negative data, thus it is meaningful to discuss their cases in comparison with that of Mo$_2$GaC for possible explanation of its *c*-axis elongation behaviour. Elastic constants calculations of Mo$_2$Ga$_{0.5}$Al$_{0.5}$C and Mo$_2$Ga$_{0.5}$In$_{0.5}$C confirmed their stable structures, as is shown in Figures **S16**,**17**, Elastic constants of Mo$_2$AlC and Mo$_2$InC are shown in Figures **S18**,**19**. The *c* axis of Mo$_2$AlC shows largest shrinkage with a value of 0.93 at 100 GPa without any anomalous behaviour. Contrarily, Mo$_2$InC shows least shrinkage with a value of 0.95 at 100 GPa. Unexpectedly, substitution of half Ga by Al and In in Mo$_2$GaC *α* phase found the abnormal phenomenon. Such as the *c* axis of Mo$_2$Ga$_{0.5}$Al$_{0.5}$C and Mo$_2$Ga$_{0.5}$In$_{0.5}$C is stiffer than their respective two end compounds within 0~60 GPa, it is Mo$_2$GaC, Mo$_2$AlC, Mo$_2$GaC, Mo$_2$InC, respectively, whereas the *c* axis of Mo$_2$Ga$_{0.5}$Al$_{0.5}$C starts to become softer than that of Mo$_2$GaC since about 60 GPa, as is shown in Figure **13**. A previous calculation for (Cr$_{0.5}$V$_{0.5}$)$_2$GeC[20] didn't find such axial compression trend.

The structural evolution of $Mo_2GaC$ $α$ and $β$ is shown in Figure **8**. Analysis to $Mo_2GaC$ and its related compounds found that the anomalous *c*-axis shrinkage behaviour of $Mo_2GaC$ seems only depends on the Mo-Mo bond, as is shown in Figure **13** and Figure **S20**, furthermore, the Mo-Ga bond shown in Figure **S7** is also smooth shrinkage under compression, suggesting the independence on the Ga atom and Mo-Ga bond, which is different with the conclusion from those of $Mo_2Ga_{0.5}Al_{0.5}C$ and $Mo_2Ga_{0.5}In_{0.5}C$, in which some complicated behaviour of *c*-axis and Mo-Mo bond contraction disappears, as is shown in Figure **13** (a), also suggesting the decisive role of the mixed Al/In, or the Ga atoms, contrarily. In detail, the introduced Al atom will induce the *c*-axis shrinkage behaviour becomes very smooth, namely, the anomaly at about 15 GPa is disappear, but the *c*-axis abnormal elongation behaviour still exists. The introduced In atom will induce the detectable but negligible fluctuation of *c*-axis shrinkage behaviour at about 15 GPa and since then the *c*-axis contracts smoothly. This interesting phenomenon could be reasonably illustrated by the relative small atomic radius discrepancy between Al and Ga 0.05~0.1 Å and large discrepancy between Ga and In 0.2~0.3 Å, respectively. In fact, the electronegativity difference between Al and Ga is also very small, the number of the charge transfer might also influences weakly to the atomic motion. The electronegativity order is Ga>In>Al.

The average atomic volume is shown in Figure **S21**, results show that the In-containing Mo"A"C has maximum volume, where Ga/Al-containing Mo"A"C has nearly same volume, which means the former case has larger buffer during the compression, as is consistent with the weak dependence of the *c* axis on the large abnormal elongation of Mo-Mo in $Mo_2InC$, as is shown in Figure **13.** Combined analysis to the *a/c*-axis (Figure **S8,9**), the Mo-C bond (Figure **S22**), and the Mo-"A" compressibilities(Figure **S7**) it is clearly seen that once we fix the Mo coordinate unchanged during the relaxation, the Mo-Mo bond will shorten smoothly without any anomaly, the *c* axis also will shorten smoothly, as is also shown in Figure **13**(c). The volume compressibility is shown in Figure **S23**. Accordingly, the additional Mo atomic motion relative to those of "A" and "X" in MAX under pressure causes the complicated *c*-axis evolution behaviour. The *c*-axis and Mo-Mo bond compressed to about 0.95 and 0.93 at 100 GPa, respectively, corresponding respectively to about 5% and 7 % reduction, far larger than that of freely moved Mo atom which only reduces about 1% at 5 GPa and then increase abnormally till to 100 GPa. The angle between Mo-Mo bond and its horizontal projection line increases with a rotation of about 1°, also far lower than the freely rotated angle with a value of about 5°. Many MAX compounds shows smooth monotonic decrease in *c*-axis despite that the "M" of MAX involves the extra motion during compression. Accordingly, the novel *c*-axis and Mo-Mo bond evolution is closely related to the Mo atomic shift.

  The current calculations observed that the $Mo_2AlC$ and $Mo_2InC$ present smooth *c*-axis shrinkage behaviour within 0~100 GPa, as is shown in Figure **13**(a), indicating that it is the Ga atom in $Mo_2GaC$ that causes the complex *c*-axis shrinkage behaviour more or less. This is indeed unambiguous according to the current data. The *c*-axis anomalous shrinkage behaviour correlates well with the Mo-Mo bond. Substitution of half Ga by Al in $Mo_2GaC$ failed to cause the *c*-axis abnormal elongation at 10~20 GPa but cause the $c$-axis abnormal elongation at 60~70 GPa shown in Figure **13**(a), respectively, resulting from the Mo-Mo abnormal elongation of $Mo_2Ga_{0.5}Al_{0.5}C$. Substitution of half Ga by In in $Mo_2GaC$ slightly decreased the $c$-axis shrinkage within 10~20 GPa but failed to cause $c$-axis abnormal elongation within 60~70 GPa, respectively. The $c$-axis slowdown shrinkage within 10~20 GPa correlates well with the Mo-Mo bond abnormal abrupt elongation.

The Mo-Mo bond length of $Mo_2AlC$, $Mo_2GaC$, and $Mo_2InC$ increase firstly and decrease subsequently under pressure, in which $Mo_2GaC$ has twice abnormal elongation but $Mo_2AlC$ and $Mo_2InC$ has only one abnormal elongation, whereas the intermetallic $Mo_2Ga_{0.5}Al_{0.5}C$ and $Mo_2Ga_{0.5}In_{0.5}C$ show one abnormal elongation, denoting all of the group members Al, Ga, In have same bonding property. Only $Mo_2GaC$ and $Mo_2Ga_{0.5}Al_{0.5}C$ have obvious $c$-axis abnormal elongation under pressure, whereas $Mo_2AlC$, $Mo_2InC$, and $Mo_2Ga_{0.5}In_{0.5}C$ display apparent $c$-axis abnormal elongation, suggesting the non-unique electronic configuration dependence, depending also on the other properties such as the slightly different atom electronegativity, the Mo-C bond strength, Mo-"A" bond strength, the $c/a$ shrinkage ratio, the rotation ratio of angle between Mo-Mo bond and its horizontal projection line, the bond population, the atomic Mulliken charge, and so on. Conclusively, this phenomenon is contributed by many factors.

Previous energy band calculations[4] for $Mo_2GaC$ observed the energy along Mo-Mo bond responds well to the Mo-Mo bond length, recent enthalpy calculation for $Mo_2GaC$ $β$ phase observes the normal contraction of $c$ axis, indicating the $c$-axis abnormal elongation of Mo$_2$GaC *α* phase at about 10 GPa might not originate directly from the structural phase transition. Moreover, the reduced *c*-axis contraction of Mo$_2$Ga$_{0.5}$In$_{0.5}$C within 10~20 GPa correlates well with the abnormal elongation of Mo-Mo bonds in Mo$_2$GaC and Mo$_2$InC, respectively.

## IV. Conclusion

The "M", "A" and "X" dependence of the phase transition pressure of the typical nanolaminate MAX ceramics compounds is conclude. A novel tetragonal *P4/mmm* is observed for the first time in MAX which is the common phase due mainly to its highly symmetric atomic arrangements at high pressure. The phase transition process completes at similar pressure is observed in this kind of compounds. The *P6$_3$/mmc* (194, *β* phase), *Cm*, *C2/m* have much lower energy at high pressure than those of the other candidate metastable phases due to their small lattice distortion relatively to the perfect *P6$_3$/mmc* (194, *α* phase). The phase transition route of these MAX compounds is generally same as the "A" layer of MAX shows a shift as a whole within *a*/*b* basal plane, sometimes also including the "M" and "X" layer shift when the lattice suffers from giant compression due to the anisotropic lattice bonding nature. Multi-phases co-existence phenomenon of Mo$_2$GaC is obtained and the *c*-axis abnormal elongation is further explained. The current careful calculation didn't repeat the experimentally obtained maximum bulk modulus of Mo$_2$GaC. Mo$_2$GaC experiences *α*→*β* phase transition, without the previously obtained intermediate Amm2. Due to the extra "M" relative shift of MAX during compression, which causes more or less effects to its neighbor atom "A" and "X", these collaboratively behaviour might induce many unusual phenomenon under pressure in $Nb_2GaN$, $Nb_2InN$, $Mo_2InN$, respectively, in particular for those metastable phases with similar energies.

Acknowledgments

Project supported by the Fundamental Research Funds for the Provincial Universities of Zhejiang (Grant No. RF-A2020002) and Natural Science Foundation of Shanxi Province (No. 201801D221035), Science Foundation of North University of China (No. XJJ201820). Ben-Yang Li is a undergraduate student applying for his bachelor degree supervised by Dr. Ze-Jin Yang.

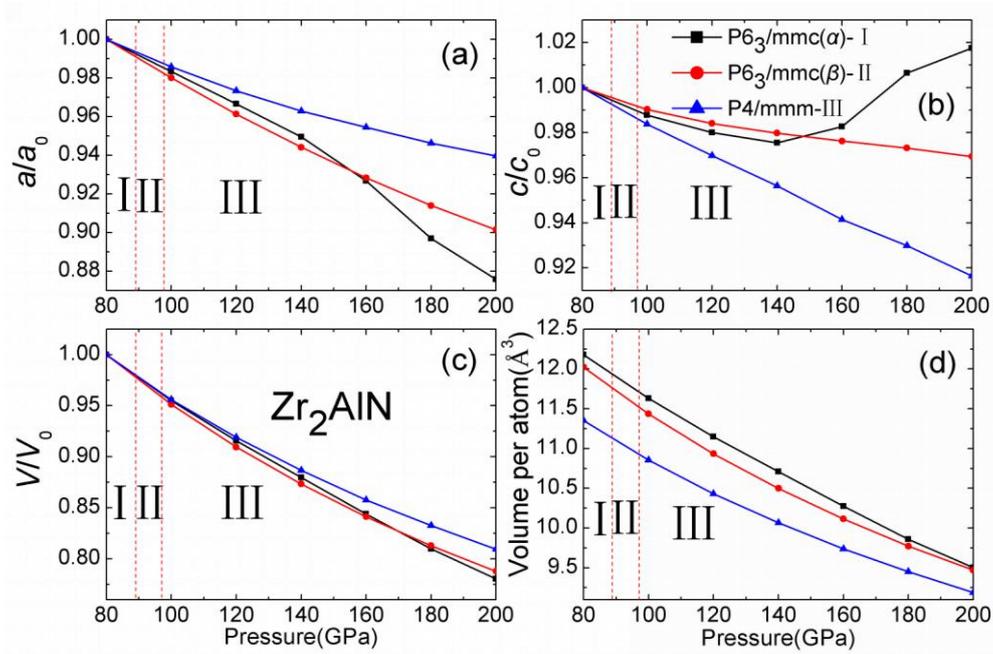

**Figure 1**. Structural evolution of Zr$_2$AlN within 80~200 GPa where phase transition occurs.

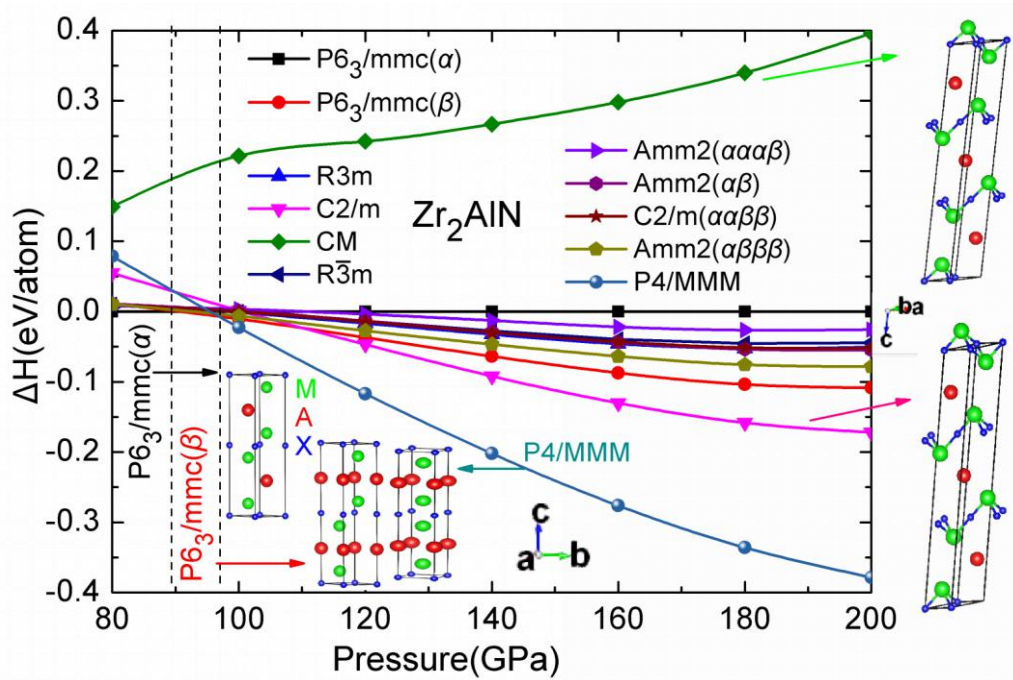

**Figure 2**. The enthalpy difference of Zr$_2$AlN within 80~200 GPa. The enthalpy of *Cm* is 0.1136 and 0.0817 eV/atom higher than that of *α* phase at 60 and 40 GPa, respectively.

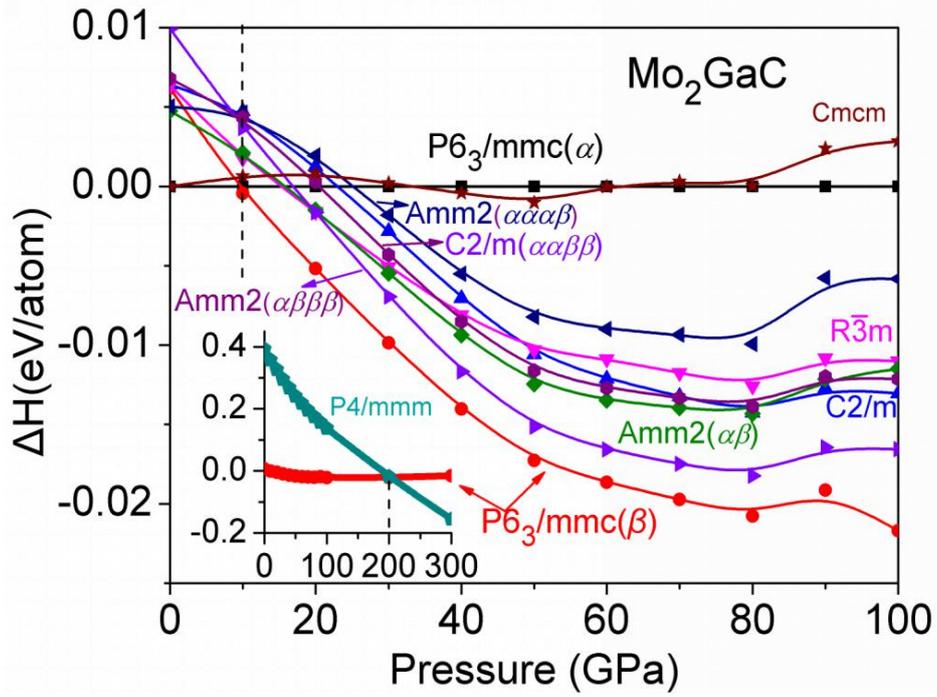

Figure 3, the phase transition of Mo₂GaC within 0~100 and 0~300 GPa for the inserted small figure.

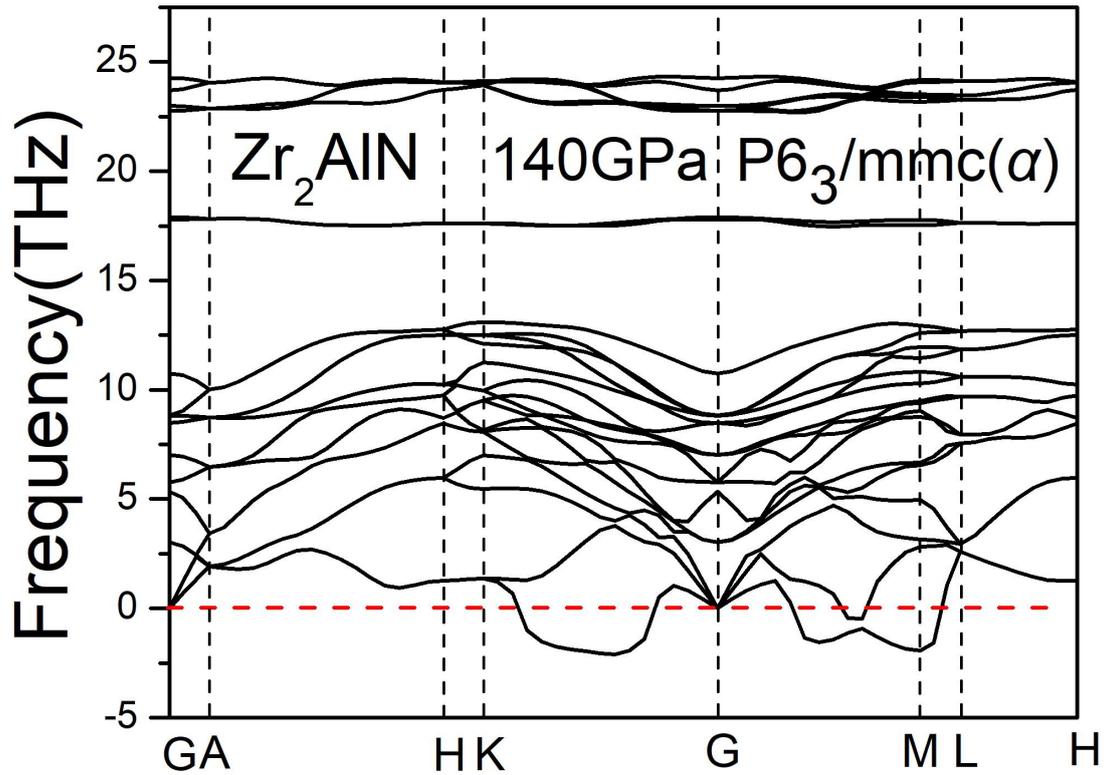

**Figure 4**. The phonon frequency of Zr₂AlN at 140 GPa.

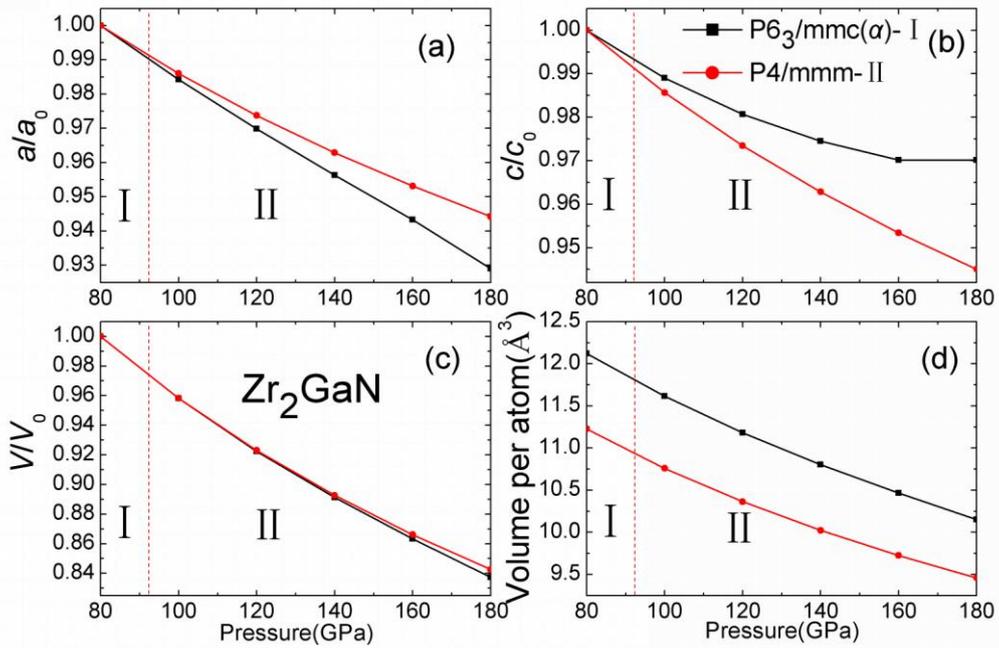

**Figure 5**. Structural evolution of $Zr_2GaN$ within 80~180 GPa.

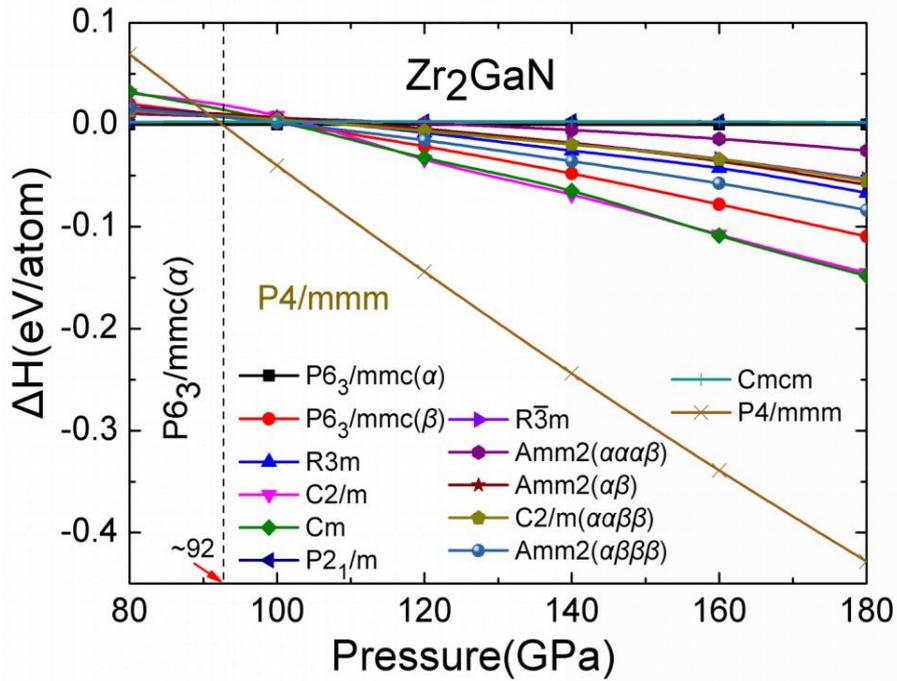

**Figure 6**. The enthalpy difference of $Zr_2GaN$ within 80~180 GPa.

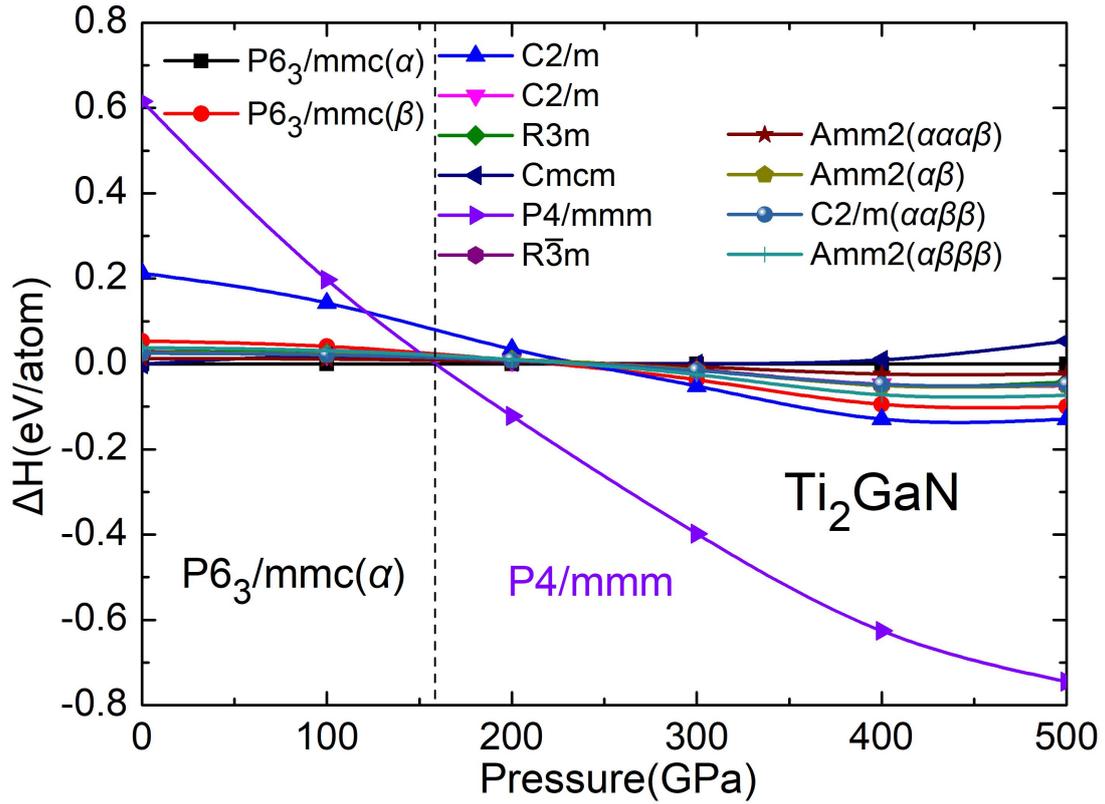

**Figure 7**. The enthalpy difference of Ti₂GaN, note that *C2/m* has two different kinds of atomic arrangements, similar with that of *P6₃/mmc*.

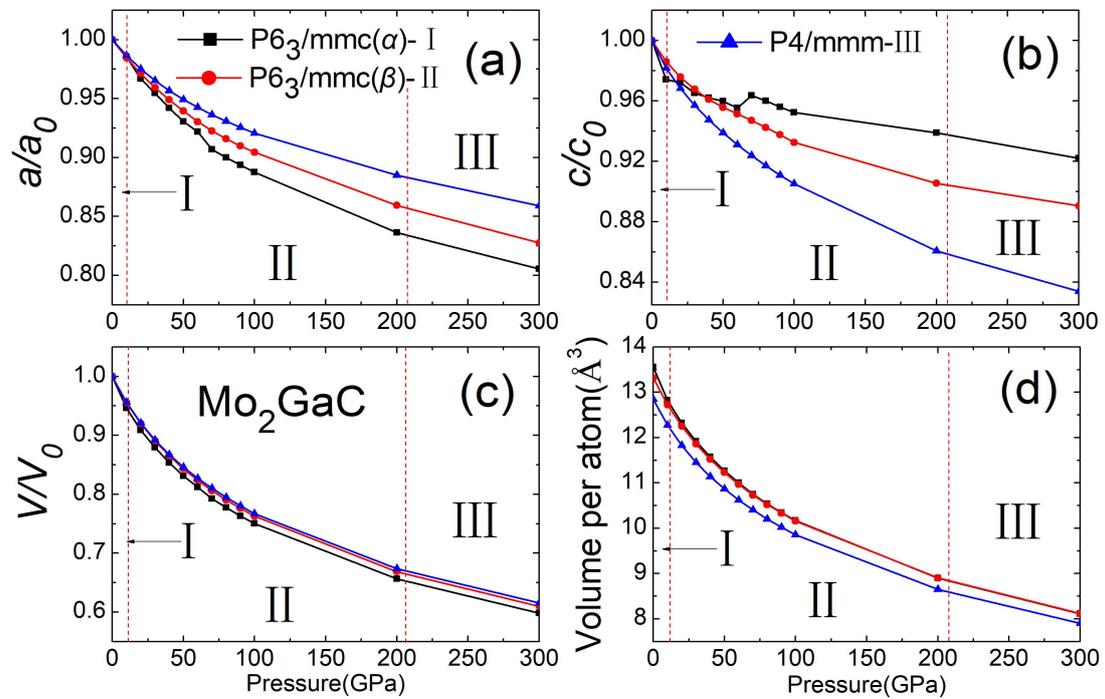

Figure 8. the structural evolution of Mo₂GaC P6₃/mmc (194, α, β phase) and P4/mmm within 0~100 GPa

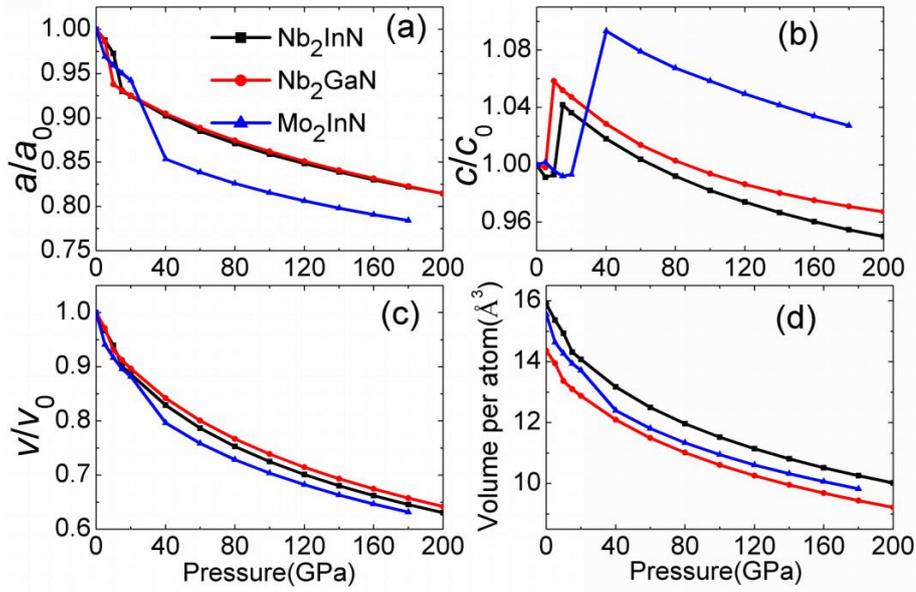

**Figure 9**. Lattice parameters of stable α phase Nb$_2$InN, Nb$_2$GaN, Mo$_2$InN within 0~200 GPa determined according to the enthalpy calculations.

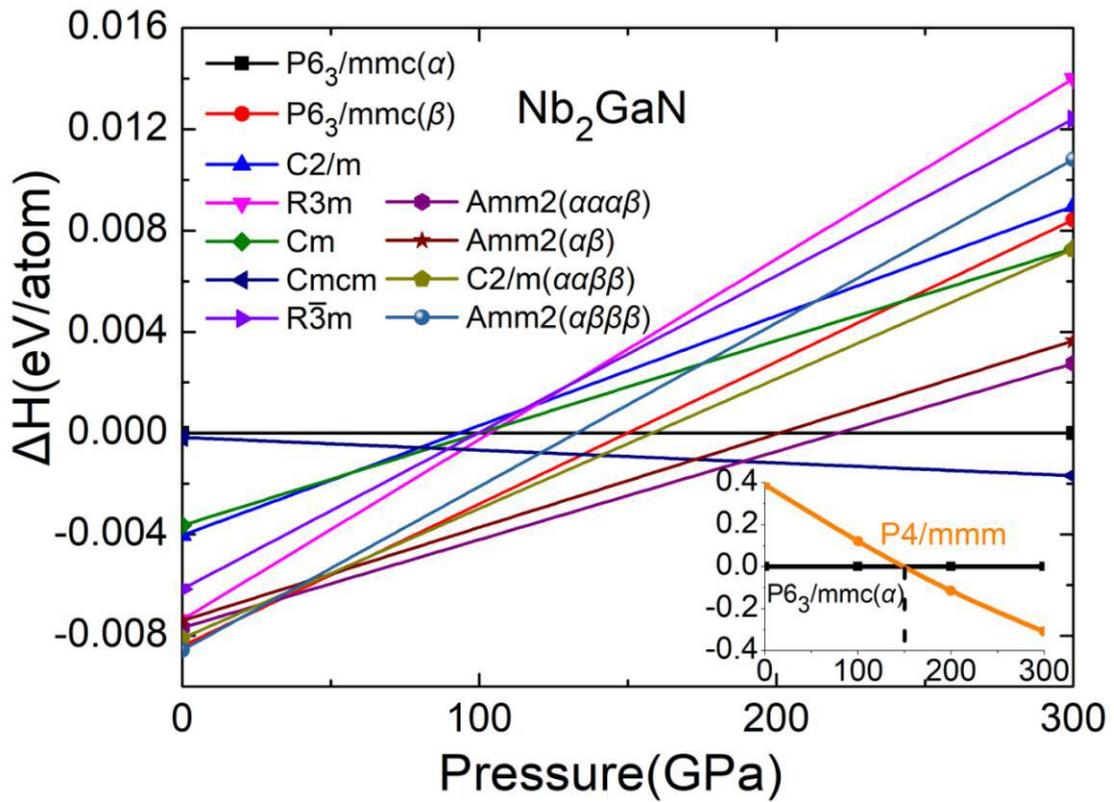

Figure 10. the enthalpy difference of Nb$_2$GaN, the inserted small figure suggests a phase transition at about 150 GPa.

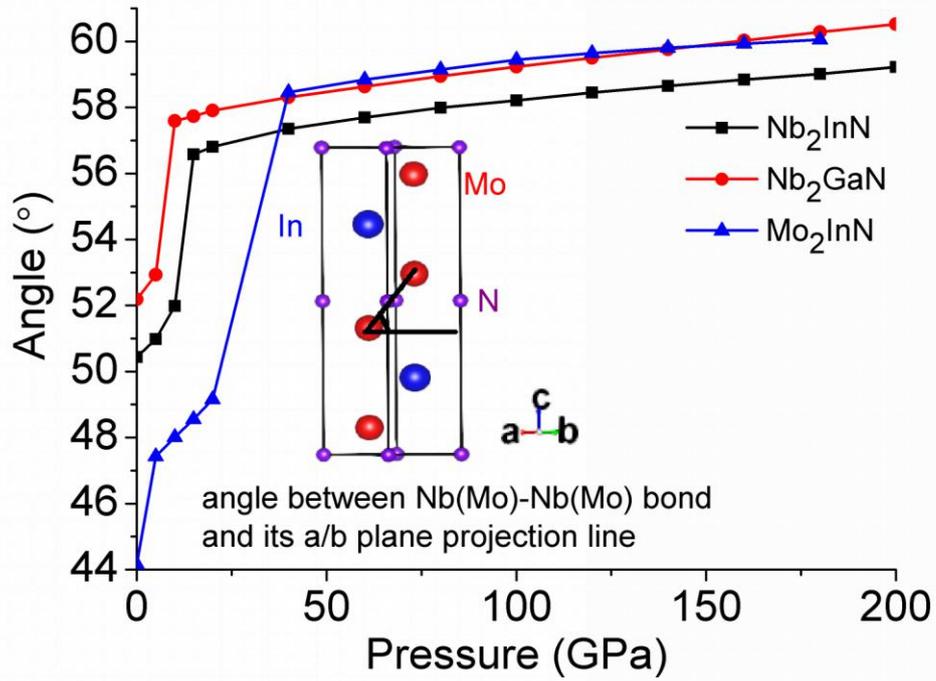

**Figure 11**. Bond angle of Nb(Mo)-Nb(Mo) bond and its *a/b* plane projection line of Nb$_2$InN, Nb$_2$GaN, Mo$_2$InN $\alpha$ phase within 0~200 GPa.

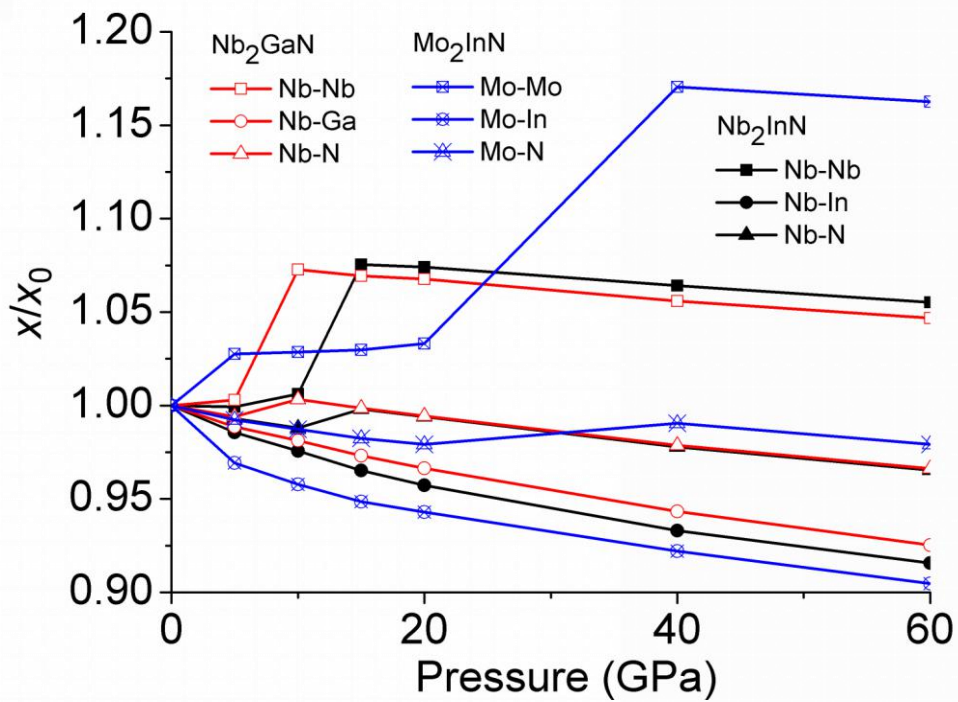

**Figure 12**. Bond length evolution of Nb$_2$InN, Nb$_2$GaN, and Mo$_2$InN within 0~60 GPa.

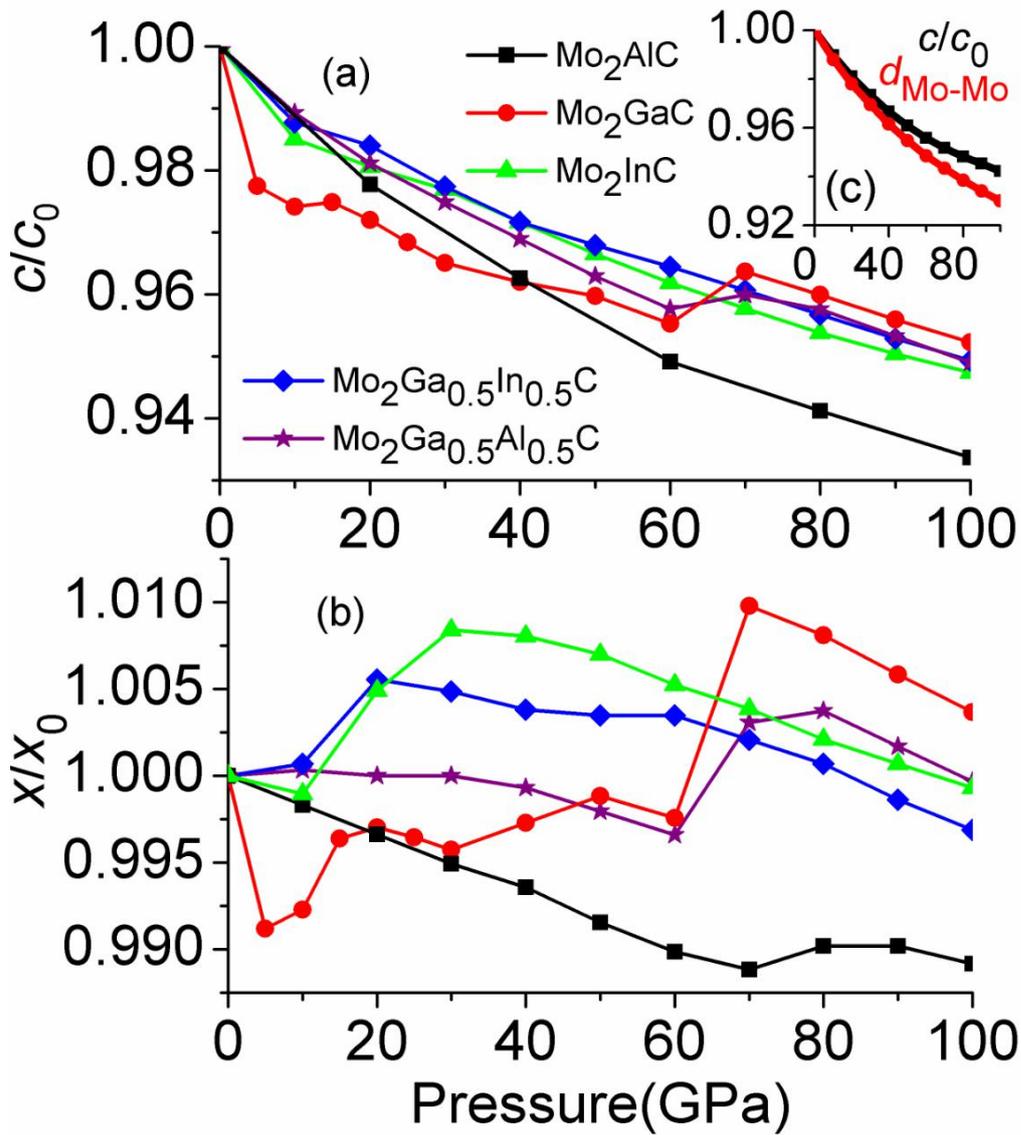

Figure 13. The *c*-axis and Mo-Mo bond shrinkage of $Mo_2AlC$, $Mo_2GaC$, $Mo_2InC$, $Mo_2Ga_{0.5}Al_{0.5}C$ and $Mo_2Ga_{0.5}In_{0.5}C$, the inserted small figure (c) is the slower *c*-axis and faster Mo-Mo bond shrinkage of $Mo_2GaC$ under the constraint of the Mo atom during the structural relaxation.